\documentclass[preprint,12pt]{elsarticle}

\usepackage{dsfont}
\usepackage{rotating}
\usepackage{multirow}

\journal{Arxiv}

\begin{document}

\begin{frontmatter}



\title{The additive property of the inconsistency degree in intertemporal decision making through the generalization of psychophysical laws}


\author[Filo]{Nat\'alia Destefano}
\ead{	nataliafm@pg.ffclrp.usp.br}
\author[Filo]{Alexandre Souto Martinez}
\ead{asmartinez@usp.br}


\address[Filo]{Faculdade de Filosofia, Ci\^encias e Letras de Ribeir\~ao Preto (FFCLRP) \\
               Universidade de S\~ao Paulo (USP) \\
               Av.~Bandeirantes, 3900 \\
               14040-901  Ribeir\~ao Preto, SP, Brazil \\
               and \\
               National Institute of Science and Technology in Complex Systems}

\begin{abstract}
Intertemporal decision making involves choices among options whose effects occur at different moments. 
These choices are influenced not only by the effect of rewards value perception at different moments, but also by the time perception effect. 
One of the main difficulties that affect standard experiments involving intertemporal choices is the simultaneity of both effects on time discounting. 
In this paper, we unify the psycophysical laws and discount value functions using the one-parameter exponential and logaritmic functions from nonextensive statistical mechanics.  
Also, we propose to measure the degree of inconsistency. 
This quantity allow us to discriminate both effects of time and value perception on discounting process and, by integration, obtain other main quantities like impulsivity and discount functions. 
\end{abstract}


\end{frontmatter}


\section{Introduction}

In contrast to Physics, Economics is based on several axioms and only in the last decades they have been extensively explored by observation. 
This may lead to limitations and deviations when standard economic models are tested empirically~\cite{bouchaud}. 
Most of the revision of these problems and the formulation of new models involve an interdisciplinary context. 

Intertemporal decision making involves choices among options whose effects occur at different moments. 
The implications of these choices on everyday activities led to the search of its underlying principles. 
Mathematical functions that fairly describe the time discount process have been suggested by experiments. 
These experiments involve humans and non-human animals and are influenced by several  factors of variability.
The consensus is that delayed rewards are discounted (or undervalued) relative to immediate rewards~\cite{green1}.

The discount process may firstly be assigned to changes in perception (evaluation) of a reward value at different moments. 
However, individuals, when forming their intertemporal preferences, may estimate time intervals in a non-objective manner~\cite{takahashi5,takahashi6,west,zauberman}.
Thus, the discount process in intertemporal choices comprises not only the effect of rewards value perception at different moments, but also the time perception effect. 

One of the main difficulties in determining discount functions from experiments is the simultaneity of both effects on time discounting.
The independent analysis of each factor is not allowed by standard experiments that directly measure these functions.

From discount functions one can obtain other quantities validated by experiments. 
The \textit{impulsivity} measures the strong preference for immediate rewards over delayed ones, even though the magnitude of the delayed reward is more advantageous. 
Also, individuals tend to prefer smaller immediate rewards in the near future (reflecting impulsivity) but tend to prefer larger later rewards in the distant future. 
This preference reversal over time is referred as \textit{inconsistency} in intertemporal choices~\cite{takahashi}. 

In this paper we unify the Weber-Fechner and Stevens psycophysical laws using the one-parameter exponential and logaritmic functions from nonextensive statistical mechanics~\cite{tsallis1}. 
This allows us to propose new general discount value functions.  
The determination of the intricate dependence between value and time perception effects in the discount function may be softened exploring their additiveness in the degree of inconsistency.
Since value and time perception are additive in the inconsistency degree, experiments may be designed to measure them independently.  
By integration of the degree of inconsistency, one obtains the impulsivity and discount functions. 
The standard experiments used in the context of intertemporal decision making need to be  reformulated for better understanding of the governing processes. 

This study is outlined as follows.
In Sec.\ref{sec:revisão} we present an overview of usual experiments and some theoretical models in intertemporal decision making.  
In Sec.\ref{sec:resultados} we describe our main findings and proposals for a new class of experiments based on the inconsistency degree.
Finally, the conclusions are presented in Sec.\ref{sec:conclusão}.  

\section{Intertemporal decision making: theory and experiments}
\label{sec:revisão}

This section presents an overview of concepts and results of the literature involving intertemporal decision making. 
We start describing the \textit{discount functions} and the standard experiments in intertemporal choices. 
We present two theoretical models that aim to describe the time discounting process observed in experiments: the exponential and the hyperbolic models. 
Next, we introduce the \textit{impulsivity} and the \textit {inconsistency}, which provide basic tools to compare exponential and hyperbolic discount models. 
Other theoretical models are also addressed. 
Finally, we describe the \textit{psychophysical laws} and the association of the so-called \textit {psychophysical effects of time perception} to the temporal discounting models.

\subsection{Discount functions}
\label{subsec:função}

Intertemporal choices refer to choices between options (rewards) whose consequences occur at different times. 
Individuals subjected to intertemporal choices face a conflict (\textit{trade off}) between the utility (or value) of an immediate reward and a delayed one. 
Consider the following examples: choosing between \$10 today or \$15 in a month; choosing to spend all earnings today or to save money for the future; deciding whether smoke or not a cigarette, to preserve health. 
In intertemporal choices, the time interval between the present time and the time when the reward is delivered is referred as \textit{delay}~\cite{read}.    

Many studies have led to a strong consensus that delayed rewards ($V$) are discounted (or undervaluated) relative to immediate rewards ($V_0$)~\cite{green1}. 
The value (or utility) of a reward $V$ decreases as the time interval till its receipt ($t$) increases. 
The non-discounted (real) value of a given reward is called \textit{objective value}. 
The value to be received immediately, which is equivalent to the receipt of $V_0$ on a specified delay, is referred as the \textit{subjective value} of the reward or \textit{indifference point}. 
The subjective value behavior of a reward as a function of delay, $V(t)$, is analyzed throughout the \textit{discount functions}. 
The shape of the discount curve is a decreasing monotonic function with null asymptotic value.           

\subsubsection{Experiments}
\label{subsec:experimentos}

Experiments, with both humans and nonhuman animals, have been conducted to determine the indifference points~\cite{takahashi1,takahashi,ainslie,reynolds,mazur,green,bickel}. 
In general, in the experiments involving delay discounting with humans, the participants choose between two monetary rewards, a smaller but immediate reward and another of greater value delivered after a given delay. 
For each delay, the experiment begins with equal values for both rewards, so that a given participant chooses the immediate reward. 
The delayed reward value is kept constant while the the immediate reward value is decreased. 
Next, the participant performs a new decision-making between immediate and delayed rewards. 
This procedure is repeated till the delayed reward is preferred to the immediate one. 
The last immediate reward value chosen, $V_d$, is described as the indifference point of the respective delay. 
To avoid a possible influence of the rewards presentation order in the experiments, the reverse procedure is also examined. 
The reversed experiment starts from the lowest value for the immediate reward, so that the delayed reward is preferred. 
The immediate reward is then increased till its first value, $V_s$, is chosen. 
The indifference point is obtained from the average between $V_d$ and $V_s$. 
The indifference points obtained for different delays are fitted and described by discount functions. 

In most experiments involving intertemporal choices hypothetical rewards are used. 
Also, the delays are not experienced by the individual during the experiment. 
This type of procedure has the advantage of being cheap and time efficient. 

To check the results validity, few studies have compared experimental data for procedures involving hypothetical and real rewards. 
In the latter, a response is randomly selected among the choices made by the participant, so that one can receive a real reward, according to value and delay chosen~\cite{kirb}. 
Up to Johnson and Bickel study~\cite{Johnson}, no experiment analyzed the same participants in both conditions (real and hypothetical). 
In their study, no significant differences have been observed between real and hypothetical procedures. 
However, one must consider that the reward values and delays used in the real experiments were smaller than those used in hypothetical procedures. 
Madden et al.~\cite{madden} analyzed the same reward values and delays for both procedures. 
In all cases, the reward objective value was \$10. 
Again, no differences were observed between both types of experiment.

Despite the results, it cannot be stated that hypothetical experiments can replace real experiments in studies involving intertemporal decision making. 
Due to the use of small rewards and delays in real experiments, further studies are necessary to extend this result to higher reward values and delays. 
Furthermore, one should consider the possible influence of experiment sequential runs.
In these runs, real and hypothetical conditions are evaluated for the same individual. 
It is suitable to perform the same procedure at least twice (sequential) for each participant. 
This does not guarantee the independence between the answers of both questionnaires.

There is also a third category of experiments on discount which consists of real-time measurements. 
These measures differ from both experiments described above because participants experience the consequences associated with their choices (rewards and delays) while completing the experiment~\cite{kirk, lane, reynolds2}.  
This type of procedure involves short delays ($<90$s) and smaller rewards($<\$0.50$) compared with other tests~\cite{reynolds1}. 
Real-time measurements can better assist researchers in determining short-term changes in delay discounting, like drug effects of addicts~\cite{donald,reynolds3,reynolds4}. 
Moreover, real-time measurements are the most appropriate tools for the analysis of discount in children. 
In this case the abstraction necessary to evaluate delays and rewards is smaller compared to other experiments. 
Despite the possible advantages of this method, its use is considerably less common than the other two experiments.
It increases expenses and time required to perform the procedure~\cite{reynolds1}. 

In addition to the varied reward type (real or hypothetical), the standard experiments of intertemporal choices allows the emergence of other possible sources of variability, such as: 
\begin{enumerate}
\item effect of presentation order (ascending or descending) of the immediate rewards~\cite{roblesa}; 
\item effect of presentation order (ascending or descending) of the delays~\cite{loew}; 
\item the signal effect, which suggests different levels of discount for losses and gains~\cite{murphy} and
\item the magnitude effect, which suggests different levels of discount for different reward values~\cite{thaler}.
\end{enumerate}

Other variability factors or ``anomalies'' have also been reported in the literature. Loewenstein and Prelec~\cite{loew2} enumerate  a set of anomalies, including the \textit{gain-loss asymmetry} and the \textit{delay-speedup asymmetry}, and proposed a model that accounts for theirs, as well as other intertemporal choice phenomena.

A less explored aspect in these experiments is the length to which participants can automatically judge the delayed rewards as uncertain. 
Even if the uncertainty about the delivery of these rewards is not taken into account  on the issues of discounting experiments. 
Patak and Reynolds~\cite{patak} used a procedure in which, immediately after the conventional delay discounting measures, participants received a form where they were questioned about their notions of uncertainty related to the delayed rewards. 
As a result, a significant correlation ($r =$ 0.55) between uncertainty and discount degree was found. 
Takahashi et al.~\cite{takahashi7} examined whether delay discoun\-ting is attributable to a decrease in the subjective probability of obtaining delayed rewards.
The results indicated that the subjective probability as a function of delay decays hyperbolically and a significant positive correlation between delay discount rate and subjective probability decay rate
was found.
However, subjective-probability discounting was not significantly correlated with delay discounting.

Delay discounting measures are designed to index the discount of delayed rewards specifically as a function of its delay. 
However, as previously described, the standard experiments for measurements of $V(t)$ are influenced by variability factors. 
These factors affect the results interpretation and consequent development of theoretical models, since it is not possible to isolate the delay as the only variable in the process. 
Therefore, these experiments need to be reviewed and reformulated for the correct interpretation of the processes involved in intertemporal decision making.

\subsubsection{Theoretical models}
\label{subsec:modelos}

Despite the difficulty of measuring $V(t)$, theoretical models have been studied to obtain mathematical functions (discount functions) that adequately describe the experimental delay discounting process. 
To start, we describe two of the main discount functions proposed: the exponential and the hyperbolic ones. 

The standard economic theory assumes rational decision makers. 
In this model, the present value of a future reward decreases exponentially~\cite{samuelson}:
\begin{equation}
V_{0n}(t)= V_0 e^{-k_0 t} \; ,
\label{exp}
\end{equation}
where $V_{0n}$ and $V_0$ correspond, respectively, to the subjective and objective reward values, and $k_0$ is the degree that an individual discounts delayed rewards. 
Higher $k_0$ values correspond to discount curves with more pronounced decay. 

In the exponential discounting model, the preference between two temporal rewards does not depend on the time the choice is made. 
If the rewards are displaced by the same time interval, the preference between them remains the same.

However, experimental results~\cite{kirb,madden,reynolds2,rodriguez,rachlin,mcker} show that the discount of a reward according to its delay is better described by a hyperbolic function~\cite{mazur}:
\begin{equation}
V_{1n}(t)= \frac{V_0}{1+k_1t} \; ,
\label{hip}
\end{equation}
where $V_{1n}$ and $V_0$ correspond to the subjective and objective reward values, respectively, and $k_1$ is a free parameter. 

In the following, we introduce some concepts used in the intertemporal choices context. 
These concepts are useful to understand the delay discounting features that lead to its better description by a hyperbolic (and not a exponential) function.  

\subsection{Impulsivity and degree of inconsistency  in intertemporal choices}
\label{subsec:impulsividade}

In intertemporal choices, \textit{Impulsivity} is defined as the strong preference for smaller, immediate rewards to greater, delayed ones~\cite{takahashi}. 
For example, suppose the following question: ``Do you prefer \$10 in a year or \$15 in a year and a week?''. 
If an individual A prefers the first option (\$10 in a year) while B prefers the second option (\$15 in a year and a week), it is said that A is more impulsive than B because A prefers a smaller, but more immediate reward, whereas B prefers to wait a longer time interval to receive a greater reward.

The term ``impulsivity'' is not restricted to the issues involving delay discounting. 
Studies have tried to verify whether discount measures assess the same process as the more traditional psychometric impulsivity measures~\cite{madden1, kirb1}. 
However, few studies have compared these two measure types and there is still a huge need of evaluating this relationship. 
Here, to avoid the various connotations attributed to ``impulsivity'', we use the definition of ``discount rate'' as a measure of impulsivity in the context of intertemporal decision making.

The \textit{discount rate} in intertemporal choices is~\cite{takahashi}:
\begin{equation}
I =-\frac{1}{V} \frac{dV}{dt} \; ,
\label{tax}
\end{equation}
the relative variation of the discount function $V$. 
The opposite behavior to impulsivity is \textit{self-control}.  

Returning to the previous example (where individual A is the more impulsive than B), consider now the following question: ``Do you prefer \$10 today or \$15 in a week?''. 
If individual B (who chose the second option in the previous question) now prefer the first option (\$10 today), his intertemporal choice is said to be dynamically inconsistent, since in both cases the same gain (\$5) is obtained in the same time interval (one week).

Experiments involving humans and nonhuman animals have shown that individuals tend to prefer smaller immediate rewards in the near future but tend to prefer larger later rewards in the distant future. 
This preference reversal over time is referred as \textit{dynamic inconsistency} in intertemporal choices~\cite{takahashi}.

Suppose that a smaller (\$7) but immediate reward is delivered at the instant $t_S$, while $t_L$ represents the instant of delivery of a larger (\$10) but delayed reward. 
The subjective value of both rewards decays as delay increases, till their respective decay curves intersect at a instant $t_E$ before both rewards delivery. If a choice is made at any time after $t_E$, the smaller immediate reward is preferred, even if their value is smaller than the other reward, reflecting impulsivity. 
However, if the choice is made at some time before $t_E$, the larger later reward is preferred, reflecting self-control.

The quantity to measure the inconsistency degree was defined by Prelec in 2004~\cite{prelec} and interpreted by Takahashi, in 2010, as the discount rate temporal variation:
\begin{equation}
\mathds{I}= \frac{dI}{dt} \; ,
\label{inc}
\end{equation}
where $I$ is the discount rate defined by Eq.~\ref{tax}. 

\subsection{Exponential and hyperbolic models}
\label{subsec:comparação}

Once the discount rate and inconsistency degree have been defined in the intertemporal choices context, we proceed with the analysis of the exponential and hyperbolic discount functions. 

For the exponential decay model (Eq.~\ref{exp}), the discount rate is constant  
\begin{equation}
I_{0n}(t)=k_0 \; ,
\label{taxe}
\end{equation}
it does not depend on the delay and $\mathds{I}_{0n}(t)=0$. 

For the hyperbolic discount model (Eq.~\ref{hip}), the discount rate is a decreasing function of $t$:
\begin{equation}
\label{taxh}
I_{1n}(t)=\frac{k_1}{1+k_1t} \; . 
\end{equation}
A reward value is strongly discounted in relatively short delays, but it is discounted at a more moderate form as the delay increases.
For this model, Eq.~\ref{inc} can be written as:
\begin{equation}
\mathds{I}_{1n}(t)= -\left[\frac{k_1}{1+k_1t}\right]^2 = -[I_{1n}(t)]^2 = \mathds{H}[I_{1n}(t)] \; ,
\label{inch}
\end{equation} 
where 
\begin{equation}
\mathds{H}(I)= -I^2 \; .
\end{equation} 
The inconsistency degree for the hyperbolic discount model is non-null and can be written as a function of the discount rate $I_{1n}$.    

Studies in intertemporal choices show that a reward discount rate decreases as the delay increases~\cite{thaler}. 
This behavior is adequately described by the hyperbolic discount model. In this case, as described above, the discount rate is a decreasing function of $t$, resulting in higher discount rates for smaller delays.

Moreover, as previously described, experiments involving humans and nonhuman animals showed a preference reversal over time. 
The exponential discount function is not able to predict this inconsistency, since $\mathds{I}_{0n}=0$. 
Assuming the same discount rate for both rewards, the discount curves of the smaller immediate reward and to the large later do not cross, so that the preference between them does not change, regardless of the moment when the decision is made. 
For the hyperbolic function, the discount rate for both rewards is inversely proportional to the delay, resulting in discount curves that intersect. 
This behavior, as described above, allows a simple interpretation for the inconsistency experimentally observed in intertemporal decision making~\cite{takahashi1}. 

\subsection{Other Models}
\label{subsec:outros}

Although the hyperbolic model describes delay discounting better than the exponential one, the experimental data are not properly adjusted with this function. 
It overestimates the subjective value for short delays, while underestimates it for large ones. 

Rachlin~\cite{rachlin1} suggested a discount function where the delay value is raised to a power $g$:  $V(t)=V_0/(1+k_1t^g)$.
This function is a particular case of the hyperbolic model (Eq.~\ref{hip}), since $V_{1n}(t^g)=V_0/(1+k_1t^g)$.
Myerson and Green proposed~\cite{myerson}:
\begin{equation}
V(t)=\frac{V_0}{(1+k_1t)^g} \; .
\label{wf}
\end{equation} 
For $g=1$, Eq.~\ref{wf} reduces to the hyperbolic model of Eq.~\ref{hip}.  
When $g<1$, a reward subjective value is more/less sensitive to changes in shorter/longer delays than in the hyperbolic model. 
Experiments~\cite{myerson,simpson} show $g\neq1$, indicating a need for a generalized model as described.
Mckerchar et al.~\cite{mcker} set the data from an experiment involving intertemporal decision making and showed the Rachlin and Myerson-Green models fit better the experimental results than the Eqs.~\ref{exp} and~\ref{hip}.

In 2006, Cajueiro~\cite{cajueiro} used a one-parameter generalization of the exponential function:
\begin{equation}
\exp_{\tilde q}(x)= \left\{\begin{array}{l}
\lim_{\tilde{q}^{'}\rightarrow\tilde{q}} (1+\tilde{q}^{'}x)^{\frac{1}{\tilde{q}^{'}}} , \ \ \ \mbox{if} \  \tilde q x\geq-1
\\
0 , \ \  \ \ \ \ \ \ \ \ \ \ \ \ \ \ \ \ \ \ \ \:\ \ \ \ \mbox{otherwise}
\end{array}
\right.
\label{qexp}
\end{equation}
where $\tilde q$ is a free parameter, $\exp_{\tilde q}(x) = e^{x}$ as $\tilde q = 0$ and $\exp_{\tilde q}(0) = 1$, for all $\tilde q$. 
The inverse of the $\tilde q $-exponential function, called as $\tilde q$-logarithm function, is defined as: 
\begin{equation}
\label{qlog}
\ln_{\tilde q}(x)=\lim_{\tilde{q}^{'}\rightarrow\tilde{q}} \frac{x^{\tilde{q}^{'}}-1}{\tilde{q}^{'}} ,
\end{equation}
where $\tilde q$ is a free parameter, $\ln_{\tilde q}(x) = \ln(x)$ as $\tilde q = 0$ and $\ln_{\tilde q}(1) = 0$.  

The generalized functions of Eqs.~\ref{qexp} and~\ref{qlog} are originated from the non-extensive thermodynamics of Tsallis~\cite{tsallis1} and have been geometrically interpreted~\cite{tiago} and can be applied in population dynamics~\cite{martinez,martinez1,2010arXiv1010.2950S} and to usual distributions in complex systems~\cite{martinez2, martinez3,takahashi2, takahashi3, takahashi4}. 

Using Eq.~\ref{qexp}, the $\tilde{q}$-generalized discount function is written as~\cite{cajueiro}: 
\begin{equation}
\label{qgen}
 V_{\tilde q n}(t) = \frac{V_0}{\exp_{\tilde q} (k_{\tilde q}t)} = \frac{V_0}{(1+\tilde{q}k_{\tilde{q}}t)^{1/\tilde{q}}} ,
\end{equation}
where $V_0$ is the objective reward value and $k_{\tilde q}$ a impulsivity parameter at $t = 0$. For $\tilde q=0$, Eq.~\ref{qgen} is equivalent to the exponential discounting function (Eq.~\ref{exp}) and for $\tilde q=1$, to the hyperbolic discount function (Eq.~\ref{hip}). 
Calling $\tilde{q}=1/g$ and $k_{\tilde{q}}= g k_1$, one retrieves the Myerson and Green model (Eq.~\ref{wf}).

For the $\tilde{q}$-generalized model, the discount rate is:  
\begin{equation}
I_{\tilde{q}n} (t)=\frac{k_{\tilde q}}{[\exp_{\tilde q}(k_{\tilde q}t)]^{\tilde q}} \; ,
\label{taxqgen}
\end{equation} 
and the time variation of this discount rate is: 
\begin{equation}
\mathds{I}_{\tilde q n} (t) = \frac{-k_{\tilde{q}}^2\tilde{q}}{(1+k_{\tilde{q}}\tilde{q}t)^2} = \tilde{q}\mathds{H}[I_{\tilde q n}(t)] \; .
\label{inqgen}
\end{equation}

This relationship can be separated into three distinct cases~\cite{takahashi3}: 
(a) $\mathds{I}_{\tilde q n}<0$ para $\tilde{q}>0$ (decreasing impulsivity);
(b) $\mathds{I}_{\tilde q n}=0$ para $\tilde{q}=0$ (exponential discounting, consistent intertemporal choices) and
(c) $\mathds{I}_{\tilde q n}>0$ para $\tilde{q}<0$ (increasing impulsivity).

The hyperbolic discount model is a particular case of (a), where $\tilde{q}=1$, and the exponential discount model corresponds to the case (b). Initial experiments~\cite{takahashi} show that, in most cases, individuals make decisions following a decreasing impulsivity pattern as a function of $t$ [case (a)].

The Cajueiro model unifies the previous models but still lacks a fundamental interpretation of $\tilde q$. 
Takahashi~\cite{takahashi1} has given a phychophysical interpretation for $\tilde{q}$ and $k_{\tilde{q}}$~\cite{takahashi}. 

\subsection{Exponential discounting with Weber-Fechner time perception}
\label{subsec:fechner}

Although  Eqs. \ref{wf} and \ref{qgen} have been suggested empirically,  Takahashi et al.~\cite{takahashi5} included the \textit{psychophysical effects of time perception} to the process of discounting in intertemporal choices.

The (invariant) ratio between stimuli that reach the possibility of being distinguished - called supraliminal stimuli - was measured by Weber in 1834. 
Later, Fechner (1860) formally expressed this invariance in the Weber's fraction ($w$), as the \textit{first psychophysical law}: 
\begin{equation}
\label{weber}
w=\frac{\Delta\phi}{\phi_p}=\frac{\phi_c-\phi_p}{\phi_p} \; ,
\end{equation}
where $\phi_c$ and $\phi_p$ are perceived stimuli in a given sensory modality, having between them a threshold $\Delta\phi\neq0$. 
The Weber's fraction is dimensionless and can be expressed as a percentage of the standard stimulus, which is specific to each sensory modality.

Possibly, there is a dependency between the ability of species discriminate stimuli of a particular sensory modality and the constancy in the expansion or contraction process of the sensation to the physical reality. 
According to Fechner, the dependency between perception and stimulus is logarithmic. 
This relationship is known as the \textit{second psychophysical law} or the \textit{Weber-Fechner law}:		
\begin{equation}
\tau (t)=a\ln(1+bt) \; ,
\label{fechner}
\end{equation}
where $\tau$ is the subjective time, and $a$ and $b$ are psychophysical parameters. 	

If one discounts delayed rewards exponentially (Eq.$\ref{exp}$), but with a subjective time perception following the Weber-Fechner law (Eq.$\ref{fechner}$), his/her time discounting is~\cite{takahashi5}:

\begin{eqnarray}
V_{0f}(t) & = & V_0 e^{-k_0 \tau}= V_0 e^{-k_0 a\ln(1+bt)} = \frac{V_0}{(1+bt)^g} \; ,
\label{ew}
\end{eqnarray} 
where $b$ and $g = k_0a$ are free parameters.

The discount rate for this case can be written as:
\begin{equation}
I_{0f}(t)=\frac{bg}{1+bt} \; .
\label{taxew}
\end{equation} 
Note that $I_{0f}$ is a decreasing function of $t$ when $b$ and $g$ are positive, resulting in preference reversal over time. 
In this case, the inconsistency degree is expressed as:
\begin{equation}
\label{incew}
\mathds{I}_{0f}(t)= \frac{-b I_{0f}(t)}{1+bt} =\mathds{F}[I_{0f}(t)], 
\end{equation} 
where 
\begin{equation}
\mathds{F}(I)= \frac{-bI}{1+bt} \; .
\end{equation}

The generalization of models from the functions $\tilde{q}$-logarithm and $\tilde q$-expo\-nential has attracted the attention of researchers in different contexts~\cite {martinez, martinez1, martinez2, takahashi2, takahashi3, takahashi4}. 
Between the difficulties of these generalizations, however, is the interpretation of $\tilde q$ in terms of the parameters describing the studied phenomena. 
For Eq.$\ref{qgen}$, using the ratios $\tilde{q}=1/(k_0a)$ and $k_{\tilde{q}}=k_0ab$, it appears that this model is mathematically equivalent to the exponential discount model with time perception of Weber-Fechner (Eq.$\ref{ew}$). 
This equivalence was described by Takahashi~\cite{takahashi1} and allows an interpretation for $\tilde{q}$ and $k_{\tilde{q}}$ from the psychophysical parameters $b$ and $k_0a$ describing the Weber-Fechner discounting~\cite{takahashi}.

\subsection{Exponential discounting with Stevens time perception}
\label{subsec:stevens}

The relationship between perception and psychophysical stimulus was also examined by Stevens~\cite{stevens}. 
For Stevens, the perception and stimulus are related by a power law, known as the \textit{third psychophysical law} or the \textit{Stevens' law}:
\begin{equation}
\tau (t)=c(1+bt)^s \; ,
\label{stevens}
\end{equation}
where $c$ and $s > 0$ are psychophysical parameters. 
If $s<1$, the subjective time decreases as $t$ increases, resulting in an overestimation of small time intervals and an underestimation of long ones. 
In contrast, when $s>1$, subjective time grows with the increase of $t$ (underestimation of small time intervals and overestimation of long ones). 

Takahashi et al.~\cite{takahashi6} were the first to proposed a time discounting model incorporating Steven's power law of time perception.
If an individual discounts delayed rewards exponentially (Eq.~\ref{exp}), but with a subjective time perception following the Stevens' law (Eq.~\ref{stevens}), its time discount is a stretched exponential function:
\begin{equation}
 V_{0s}(t) = e^{-k_p(1+bt)^s} \; ,
\label{es}
\end{equation}
where $k_p=k_0c$ and $s$ are free parameters. For this model, the discount rate is: 
\begin{equation}
I_{0s} (t)= k_p b s(1+bt)^{s-1} \; .
\label{taxes}
\end{equation}
Note that $I_{0s}$ is a decreasing function of $t$, with $s<1$ and $k_p>0$. The inconsistency degree is:
\begin{eqnarray}
\nonumber
\mathds{I}_{0s} (t) & = & k_psb^2(s-1)(1+bt)^{s-2}=\frac{b(s-1)}{1+bt}I_{0s} \\ 
                              & = & (1-s)\mathds{F}[I_{0s}(t)] \; .
\label{inces}
\end{eqnarray}

Takahashi et al.~\cite{takahashi1} performed experiments involving intertemporal choices with 26 volunteer students. 
Their parameters were estimated from the intertemporal choice equations involving models without time perception effects (exponential discounting and hyperbolic discounting) and models that include this effect (Weber-Fechner discounting and Stevens discounting). 
The Weber-Fechner discount model (Eq.~\ref{ew}) best fitted the experimental values. 
This result is in agreement with other similar studies~\cite{green1}. 
Nevertheless, the Stevens' discount model (Eq.~\ref{es}), given by the stretched exponential function, fitted the data better than the hyperbolic discount model (Eq.~\ref{hip}). 
The exponential discount model (Eq.~\ref{exp}), which reflects time consistency, was the worst function to describe the experimental results. 

\section{Results}
\label{sec:resultados}

In the following we show two important results. 
The first is a particular and unprecedented unification of the psychophysical laws of Weber-Fechner and Stevens, obtained by using the $\tilde{q}$-logarithm function. 
Next we show that the inconsistency degree ($\mathds{I}$) allows independence between the effects of value and time perception. 
Finally, we suggest the need of a new class of experiments to analyze the discount process in intertemporal choice.

\subsection{Unification of the psychophysical laws}
\label{subsec:unificação}

A successful attempt to unify the psychophysical laws of Weber-Fechner and Stevens was made by Wong and Norwich in 1997~\cite{wong}. 
Here, we present an unprecedented approach of this unification, using the $\tilde q$-logarithm function.
From the Stevens'law (Eq.~\ref{stevens}), we write: $(\tau-c)/s = [c(1+bt)^s-c]/s = c [(1+bt)^s-1]/s$. 
Using this relation and the definition of Eq.~\ref{qlog} we rewrite Eq.~\ref{stevens} as:  
\begin{equation}
\tau(t)=a\ln_s(1+bt)+c
\label{unific}
\end{equation}
where $a=cs$ and $c=\tau_0$ is interpreted as a basal sensitivity. Let us take the constant $a$ as an independent quantity  of $s$ and $c$. 
For simplicity, we take null basal sensitivity ($c = 0$). 
This equation corresponds to a new unification of the psychophysical laws of Weber-Fechner and Stevens, where for the particular case $s=0$, we retrieve the Weber-Fechner law and, for otherwise, the Stevens' law.

The unified form of the psychophysical laws (Eq.~\ref{unific}) can be used in association with the exponential (Eq.~\ref{exp}) and the hyperbolic (Eq.~\ref{hip}) discount models. 
The functions:  
\begin{eqnarray}
V_{0u}(t) & = & \frac{V_0}{\exp[k_0a\ln_s(1+bt)]}
\label{eu} \\
I_{0u}(t)  & =  & \frac{- d  \ln V_{0u}}{dt} = \frac{I_0^{(0)} [s \ln_s(1+bt)+1]}{1+bt}
\label{taxeu} \\
\nonumber
\mathds{I}_{0u}(t) & = & \frac{ d I_{0u}}{dt}  = \frac{b(s-1)}{1+bt}I_{0u}(t) \\
                             & = & (1-s)\mathds{F}[I_{0u}(t)]
\label{inceu}
\end{eqnarray}
with $I_0^{(0)} = k_0 a b $, refer to exponential discount and for hyperbolic dicounting:  
\begin{eqnarray}
V_{1u}(t) & = & \frac{V_0}{1+k_1a\ln_s(1+bt)}
\label{hu} \\
I_{1u}(t) & = & \frac{- d  \ln V_{0u}}{dt} = \frac{I_0^{(1)} [s\ln_s(1+bt)+1]}{[1+k_1a\ln_s(1+bt)](1+bt)}
\label{taxhu} \\
\nonumber
\mathds{I}_{1u}(t) & = & \frac{ d I_{0u}}{dt}  =  -I_{1u}^2-\frac{b}{1+bt}I_{1u}(t) \\
                             & = & \mathds{H}(I_{1u})+ (1-s)\mathds{F}[I_{1u}(t)] \ ; ,
\label{inchu}
\end{eqnarray}
with $I_0^{(1)} = k_1 a b $. 

Furthermore, from Eqs.~\ref{unific} and~\ref{qgen} one obtains a generalized function for the discount process involving the time perception psychophysics:
\begin{equation}
V_{\tilde q u}(t)=\frac{V_0}{\exp _{\tilde q}[k_{\tilde q}a\ln_s(1+bt)]}
\label{qu}
\end{equation}
The expressions of impulsivity and inconsistency for this generalized model are written as:
\begin{equation}
I_{\tilde q u} (t)=\frac{I_0^{(\tilde q)} [s\ln_s(1+bt)+1]}{[\exp_{\tilde q}(k_{\tilde q}a\ln_s(1+bt))]^{\tilde q}(1+bt)} 
\label{taxqu}
\end{equation}
where $I_0^{(\tilde q)}=I_{\tilde q u}(0)=k_{\tilde q}ab$  and 
\begin{eqnarray}
\nonumber
\mathds{I}_{\tilde q u} (t) & = & -\tilde q I_{\tilde q u}^2(t) + \frac{(s-1)b}{1+bt}I_{\tilde q u}(t) \\
                                        & = &\tilde q \mathds H[I_{\tilde q u}(t)] + (1-s)\mathds F[I_{\tilde q u}(t)] \; .
\label{incqu}
\end{eqnarray}

From Eq.~\ref{qu}, one obtains particular models from the variation of parameters $\tilde q$ and $s$. 
For $s=0$, which corresponds to the Weber-Fechner time perception, we write:
\begin{eqnarray}
V_{\tilde q f} (t) & = & \frac{V_0}{\exp _{\tilde q}[k_{\tilde q}a\ln(1+bt)]}
\label{qw} \\
I_{\tilde q f} (t) & = & \frac{I_0^{(\tilde q)}}{(1+bt)[\exp_{\tilde q}(k_{\tilde q}a\ln(1+bt))]^{\tilde q}}	
\label{taxqw} \\ 
\nonumber
\mathds{I}_{\tilde q f} (t) & = & -\tilde q I_{\tilde q f}^2(t) -\frac{b}{1+bt}I_{\tilde q f}(t)  \\
                                       & = & \tilde q \mathds H[I_{\tilde q f}(t)] + \mathds F[I_{\tilde q f}(t)]
\label{incqw}
\end{eqnarray}

For $\tilde q =0$, which corresponds to the exponential discount model, Eqs. $\ref{qw}$, $\ref{taxqw}$ and $\ref{incqw}$ correspond to Eqs. $\ref{ew}$, $\ref{taxew}$ and $\ref{incew}$, respectively. 
For $\tilde q =1$, we obtain:  

\begin{eqnarray}
V_{1f}(t) & = & \frac{V_0}{1+k_1a\ln(1+bt)}
\label{hw} \\
I_{1f} (t) & = & \frac{I_0^{(1)}}{(1+bt)[1+k_1a\ln(1+bt)]}
\label{taxhw} \\
\nonumber
\mathds{I}_{1f}(t) & = & -I_{1f}^2(t) - \frac{b}{1+bt}I_{1f}(t) \\ 
                            & = & \mathds H[I_{1f}(t)]+ \mathds F[I_{1f}(t)] 
\label{inchw} \; .
\end{eqnarray}

\subsection{Additive property of inconsistency}
\label{subsec:adição}

Table \ref{tabela:geral} summarizes the possible associations between the discount models without time perception (exponential discounting, hyperbolic discounting and  $\tilde q $-generalized discounting) and the psychophysical laws (Weber-Fechner law and Stevens'law), including the unified form we have proposed.

\begin{sidewaystable}[!htbp]
	
	\caption{Discount models in intertemporal choices. 
	              Subindexes 0, 1 and $\tilde q$ represent the: exponential, hyperbolic and $ \tilde q$-generalized  discount models, respectively. 
	              Subindexes $f$,$s$ and $u$ represent the: Weber-Fechner, Stevens'  and unified laws, respectively. 
	              Note that, from all associations, only five cases were considered in the literature so far: the three models with no perception effect (exponential, hyperbolic and $\tilde q$-generalized) and the association of the exponential discounting with Weber-Fechner and Stevens time perception. 
}
		
\begin{center}	
${\mathds H(I) = -I^2}\ ;\ {\mathds F(I) = -bI/(1+bt)}$
\end{center}

		\begin{tabular}{  c  c | c | c | c | }

			 \cline{3-3}	\cline{4-4} \cline{5-5}

			    &      \multirow{2}{*}{}	&	\multirow{2}{*}{\footnotesize{Exponential discounting (0)}}	&	\multirow{2}{*}{\footnotesize{Hyperbolic discounting (1)}}&\multirow{2}{*}{\footnotesize{$\tilde q$-generalized discounting ($\tilde{q}$)}}	\\
				& 						&										& 																&										\\
			\hline

	\multicolumn{1}{|c|}{	\multirow{8}{*}{\begin{sideways}$V(t)$\end{sideways}}} &\multicolumn{1}{|c|}{\multirow{2}{*}{\footnotesize{No perception ($n$)}}}	& \multirow{2}{*}{\footnotesize Eq.~\ref{exp}}  	& \multirow{2}{*}{\footnotesize Eq.~\ref{hip}}	& \multirow{2}{*}{\footnotesize Eq.~\ref{qgen}}	\\
				\multicolumn{1}{|c|}{}		&										&				\tiny {(rational agent)}					&																	&			\\	\cline{2-2} \cline{3-3}	\cline{4-4}	\cline{5-5}
			
		\multicolumn{1}{|c|}{}	&   \multicolumn{1}{|c|}{\multirow{2}{*}{\footnotesize{Weber-Fechner ($f$)}}}	& \multirow{2}{*}{\footnotesize Eq.~\ref{ew}} 	& \multirow{2}{*}{\footnotesize Eq.~\ref{hw}}	& \multirow{2}{*}{\footnotesize Eq.~\ref{qw}}	\\  	
			\multicolumn{1}{|c|}{}				&		\multicolumn{1}{|c|}{}													&											&																	&											\\	 		\cline{2-2} \cline{3-3}	\cline{4-4}	\cline{5-5}		
																						
	\multicolumn{1}{|c|}{}	&	\multicolumn{1}{|c|}{\multirow{2}{*}{\footnotesize{Stevens ($s$)}}}	& \multirow{2}{*}{\footnotesize Eq.~\ref{eu}}  & \multirow{2}{*}{\footnotesize Eq.~\ref{hu}}	& \multirow{2}{*}{\footnotesize Eq.~\ref{qu}} \\	
			\multicolumn{1}{|c|}{}	&							\multicolumn{1}{|c|}{}											&											&																	&	    							\\ \cline{2-2} \cline{3-3}	\cline{4-4}	\cline{5-5}

	\multicolumn{1}{|c|}{}		&\multicolumn{1}{|c|}{ \footnotesize{Unified form of}} 									& \multirow{2}{*}{\footnotesize Eq.~\ref{eu}}	& \multirow{2}{*}{\footnotesize Eq.~\ref{hu} }			&	\multirow{2}{*}{\footnotesize Eq.~\ref{qu}}	\\
	\multicolumn{1}{|c|}{}	& 	\multicolumn{1}{|c|}{\footnotesize{psychophysical laws ($u$)}}										&											&										&		       															
	\\		\cline{2-2} \cline{3-3}	\cline{4-4}	\cline{5-5}
			\hline

\multicolumn{1}{|c|}{	\multirow{8}{*}{\begin{sideways}$I(t)=-(dV/dt)/V$\end{sideways}}} & \multicolumn{1}{|c|}{\multirow{2}{*}{\footnotesize{No perception ($n$)}}}	& \multirow{2}{*}{\footnotesize Eq.~\ref{taxe}} 	& \multirow{2}{*}{\footnotesize Eq.~\ref{taxh}}	& \multirow{2}{*}{\footnotesize Eq.~\ref{taxqgen}}	\\
						\multicolumn{1}{|c|}{}		&	    											&			\tiny{(rational agent)}								&																	&											\\	\cline{2-2} \cline{3-3}	\cline{4-4}	\cline{5-5}		
			
	\multicolumn{1}{|c|}{}	&	\multicolumn{1}{|c|}{\multirow{2}{*}{\footnotesize{Weber-Fechner ($f$)}}}	& \multirow{2}{*}{\footnotesize Eq.~\ref{taxew}} 	& \multirow{2}{*}{\footnotesize Eq.~\ref{taxhw}} &	 \multirow{2}{*}{\footnotesize Eq.~\ref{taxqw}}	\\
				\multicolumn{1}{|c|}{}						&		\multicolumn{1}{|c|}{}										&											&																	&											\\					\cline{2-2} \cline{3-3}	\cline{4-4}	\cline{5-5}		
																						
	\multicolumn{1}{|c|}{}&		\multicolumn{1}{|c|}{\multirow{2}{*}{\footnotesize{Stevens ($s$)}}}	& \multirow{2}{*}{\footnotesize Eq.~\ref{taxeu}}  & \multirow{2}{*}{\footnotesize Eq.~\ref{taxhu}}	& \multirow{2}{*}{\footnotesize Eq.~\ref{taxqu}}	\\
		\multicolumn{1}{|c|}{}				&					\multicolumn{1}{|c|}{}											&											&																	&											\\
\cline{2-2} \cline{3-3}	\cline{4-4}	\cline{5-5}		
			
	\multicolumn{1}{|c|}{}	&\multicolumn{1}{|c|}{	\footnotesize{Unified form of}} 					& \multirow{2}{*}{\footnotesize Eq.~\ref{taxeu}}	& \multirow{2}{*}{\footnotesize Eq.~\ref{taxhu}}							&	\multirow{2}{*}{\footnotesize Eq.~\ref{taxqu}}	\\
\multicolumn{1}{|c|}{}	&	\multicolumn{1}{|c|}{	\footnotesize{psychophysical laws ($u$)}}				&											&																	&											\\

			\hline

				\multicolumn{1}{|c|}{\multirow{8}{*}{\begin{sideways}$\mathds{I}(t)=dI/dt$\end{sideways}}} & \multicolumn{1}{|c|}{\multirow{2}{*}{\footnotesize{No perception ($n$)}}}	& \multirow{2}{*}{\footnotesize $\mathds{I}_{0n}=0$} 	& \multirow{2}{*}{\footnotesize $\mathds{I}_{1n}=\mathds H(I_{1n})$}	& \multirow{2}{*}{\footnotesize $\mathds{I}_{\tilde q n}=\tilde {q} \mathds H(I_{\tilde q n})$}	\\
						\multicolumn{1}{|c|}{}		&  				& \tiny{(rational agent)}											&																	&											\\				\cline{2-2} \cline{3-3}	\cline{4-4}	\cline{5-5}		
			
	\multicolumn{1}{|c|}{}	&	\multicolumn{1}{|c|}{\multirow{2}{*}{\footnotesize{Weber-Fechner ($f$)}}}	& \multirow{2}{*}{\footnotesize$\mathds{I}_{0f}=\mathds F(I_{0f})$} 	& \multirow{2}{*}{\footnotesize$\mathds{I}_{1f}=\mathds H(I_{1f})+ \mathds F(I_{1f})$}	& \multirow{2}{*}{\footnotesize$\mathds{I}_{\tilde q f}=\tilde q \mathds H(I_{\tilde q f})+ \mathds F(I_{\tilde q f})$}	\\
							\multicolumn{1}{|c|}{}						&		\multicolumn{1}{|c|}{}							&											&																	&											\\					\cline{2-2} \cline{3-3}	\cline{4-4}	\cline{5-5}		
																						
	\multicolumn{1}{|c|}{}	&	\multicolumn{1}{|c|}{\multirow{2}{*}{\footnotesize{Stevens ($s$)}}}	& \multirow{2}{*}{\footnotesize$\mathds{I}_{0s}=(1-s)\mathds F(I_{0s})$}  & \multirow{2}{*}{\footnotesize$\mathds{I}_{1s}=\mathds H(I_{1s})+ (1-s)\mathds F(I_{1s})$}	& \multirow{2}{*}{\footnotesize$\mathds{I}_{\tilde q s}=\tilde q \mathds H(I_{\tilde q s})+ (1-s)\mathds F(I_{\tilde q s})$}	\\
	\multicolumn{1}{|c|}{}												&	\multicolumn{1}{|c|}{}								&											&																	&											\\
		\cline{2-2} \cline{3-3}	\cline{4-4}	\cline{5-5}		
			
\multicolumn{1}{|c|}{}	&	\multicolumn{1}{|c|}{	\footnotesize{Unified form of}}				& \multirow{2}{*}{\footnotesize$\mathds{I}_{0u}=(1-s) \mathds F(I_{0u})$}  & \multirow{2}{*}{\footnotesize$\mathds{I}_{1u}=\mathds H(I_{1u}) + (1-s) \mathds F(I_{1u})$}	& \multirow{2}{*}{\footnotesize$\mathds{I}_{\tilde q u}=\tilde q \mathds H(I_{\tilde q u}) + (1-s)\mathds F(I_{\tilde q u})$}	\\
\multicolumn{1}{|c|}{}		&	\multicolumn{1}{|c|}{\footnotesize{psychophysical laws ($u$)}}												&											&																	&											\\

			\hline
		\end{tabular}
	\label{tabela:geral}			
		
\end{sidewaystable}
 
As a consequence of the results involving intertemporal choices, the discount process, comprises not only the effect of rewards value perception at different moments, but also the time perception effect. 
Even though, these effects act simultaneously on the discount process and they do not allow an independent analysis of the influence of each one in the discount function form. 
Our aim is to show that the analysis of the discount process by measuring the inconsistency degree ($\mathds{I}$) can provide a solution for the dependency of value and time perceptions.

As the main result, we find that, for each association of value and time perception effects, it is possible to dissociate the functions that describe the inconsistency in two parts: one describing the value perception effect and the other the time perception one. 
Moreover, the total value of the inconsistency degree is obtained from the sum of these two terms.
For a better understanding, consider Table \ref{tabela:geral}.
Its first column relates the exponential discounting model with the psychophysical effects of time perception. 
For these cases, the inconsistency degree reflects only the contribution of the time perception effect, since the inconsistency degree for the exponential model is null.
Similarly, the first line of the inconsistency degree expressions reflects only the value perception effect. 
For other associations, the total inconsistency degree is composed of the sum of value and time effect components.
For instance, the hyperbolic discounting model in association with the Weber-Fechner time perception results in a total degree of inconsistency ($\mathds I_{1f}$) composed by the sum of the hyperbolic discounting term ($\mathds H (I_{1f})$) and the Weber-Fechner time perception term ($\mathds F (I_{1f})$).
Both components can be written as a function of the discount rate ($I_{1f}$).   
 
This finding shows that the standard experiments of discounting in intertemporal choices need to be reformulated. 
Since the quantity that allows additivity between both effects is the inconsistency degree, its direct determination favors the understanding of the determinants in discount process and their respective contributions. 
The $V$ and $I$ expressions can then be obtained by successive integration of $\mathds{I}$.

\section{Conclusions}
\label{sec:conclusão}

The neoclassical economic theory assumes rational decision-makers where the discount of real rewards in time is characterized by an exponential decay model. 
Experiments involving humans and non-human animals show that this function cannot adequately describe the results for the discount process in intertemporal choices. 
This process is better described by hyperbolic discount models. 
The discount process in intertemporal choices involves not only the effect of rewards value perception at different moments, but also the time perception effect. 

Time discounting measures are designed to index the delayed rewards discount specifically as a function of the delay to their delivery. 
The standard experiments are affected by several factors of variability. 
These factors difficult the data analysis and consequent development of theoretical models, since it is not possible to isolate the delay as the unique variable in the discount process. 
Furthermore, the concurrent action of value and time perception effects does not allow an independent analysis of the influence of each factor in determining discount functions. 
For better understanding the processes involved in intertemporal decision making the traditional experiments need to be reviewed and reformulated.

Here, we have unified the psicophysical laws and proposed a very general and unified model for the discount process. 
With this general model, we show that it is possible to dissociate the inconsistency degree $\mathds{I}$ in two independent parts of percepction: one for value and the other for time. 
Thus, the direct analysis of the inconsistency degree is the natural measure that favors the interpretation of the discount phenomenon. 
The discount functions and rates can be obtained by successive integration of $\mathds{I}$.

The association of the time perception psychophysical effects to the decision making violates the rational agent assumption based on the classic economic model.

\section*{Ackowledgements}

N. D. acknowledges support from FAPESP (2009/17733-5).
A.S.M. acknowledges support from CNPq (303990/2007-4).


\begin{thebibliography}{10}
\expandafter\ifx\csname url\endcsname\relax
  \def\url#1{\texttt{#1}}\fi
\expandafter\ifx\csname urlprefix\endcsname\relax\def\urlprefix{URL }\fi

\bibitem{bouchaud}
J.-P. Bouchaud, Economics needs a scientific revolution, Nature 455 (2008)
  1181.

\bibitem{green1}
L.~Green, J.~Myerson, A discounting framework for choice with delayed and
  probabilistic rewards, Psychological Bulletin 130~(5) (2004) 769--792.

\bibitem{takahashi5}
T.~Takahashi, Loss of self-control in intertemporal choice may be attributable
  to logarithmic time-perception, Medical Hypotheses 65~(4) (2005) 691--693.

\bibitem{takahashi6}
T.~Takahashi, Time-estimation error following weber–fechner law may explain
  subadditive time-discounting, Medical Hypotheses 67 (2006) 1372--1374.

\bibitem{west}
B.~J. West, P.~Grigolini, A psychophysical model of decision making, Physica A
  389 (2010) 3580--3587.

\bibitem{zauberman}
G.~Zauberman, B.~K. Kim, S.~A. Malkoc, J.~R. Bettman, Discounting time and time
  discounting: Subjective time perception and intertemporal preferences,
  Journal of Marketing Research XLVI (2009) 543–556.

\bibitem{takahashi}
T.~Takahashi, H.~Oono, M.~H.~B. Radford, Empirical estimation of consistency
  parameter in intertemporal choicebased on {T}sallis' statistics, Physica A
  381 (2007) 338--342.

\bibitem{tsallis1}
C.~Tsallis, What are the numbers that experiments provide?, Química Nova 17~(6)
  (1994) 468--471.

\bibitem{read}
D.~Read, Is time discounting hyperbolic or subadditive?, Journal of Risk and
  Uncertainty 23~(1) (2001) 5--32.

\bibitem{takahashi1}
T.~Takahashi, H.~Oono, M.~H.~B. Radford, Psychophysics of time perception and
  intertemporal choice models, Physica A 387 (2008) 2066--2074.

\bibitem{ainslie}
G.~W. Ainslib, Impulse control in pigeons, Journal of the Experimental Analysis
  of Behavior 21~(3) (1974) 485--489.

\bibitem{reynolds}
B.~Reynolds, H.~de~Wit, J.~B. Richards, Delay of gratification and delay
  discounting in rats, Behavioural Processes 59 (2002) 157--168.

\bibitem{mazur}
J.~E. Mazur, D.~R. Biondi, Delay-amount tradeoffs in choices by pigeons and
  rats: hyperbolic versus exponential discounting, Journal of the Experimental
  Analysis of Behavior 91~(2) (2009) 197--211.

\bibitem{green}
L.~Green, A.~F. Fry, J.~Myerson, Discounting of delayed rewards: a life-span
  comparison, Psychological Science 5 (1994) 33--36.

\bibitem{bickel}
W.~K. Bickel, A.~L. Odum, G.~J. Madden, Impulsivity and cigarette smoking:
  delay discounting in current, never, and ex-smokers, Psychopharmacology 146
  (1999) 447--454.

\bibitem{kirb}
K.~N. Kirby, Bidding on the future: Evidence against normative discounting of
  delayed rewards, Journal of Experimental Psychology: General 126~(1) (1997)
  54--70.

\bibitem{Johnson}
M.~W. Johnson, W.~K. Bickel, Within-subject comparison of real and hypothetical
  money rewards in delay discounting, Journal of the Experimental Analysis of
  behavior 2002 77~(2) (2002) 129--146.

\bibitem{madden}
G.~J. Madden, A.~M. Begotka, B.~R. Raiff, L.~L. Kastern, Delay discounting of
  real and hypothetical rewards, Experimental and Clinical Psychopharmacology
  11~(2) (2003) 139--145.

\bibitem{kirk}
J.~M. Kirk, A.~W. Logue, Effects of deprivation level on humans' self-control
  for food reinforcers, Appetite 28 (1997) 215--226.

\bibitem{lane}
S.~D. Lane, D.~R. Cherek, C.~J. Pietras, O.~V. Tcheremissine, Measurement of
  delay discounting using trial-by-trial consequences, Behavioural Processes 64
  (2003) 287--303.

\bibitem{reynolds2}
B.~Reynolds, R.~Schiffbauer, Measuring state changes in human delay
  discounting: an experiential discounting task, Behavioural Processes 67
  (2004) 343–356 67 (2004) 343--356.

\bibitem{reynolds1}
B.~Reynolds, A review of delay-discounting research with humans: relations to
  drug use and gambling, Behavioural Pharmacology 17~(8) (2006) 651–667.

\bibitem{donald}
J.~McDonald, L.~Schleifer, J.~B. Richards, H.~de~Wit, Effects of {T}{H}{C} on
  behavioral measures of impulsivity in humans, Neuropsychopharmacology 28
  (2003) 1356--1365.

\bibitem{reynolds3}
B.~Reynolds, J.~B. Richards, M.~Dassingera, H.~de~Wita, Therapeutic doses of
  diazepam do not alter impulsive behavior in humans, Pharmacology,
  Biochemistry and Behavior 79 (2004) 17--24.

\bibitem{reynolds4}
B.~Reynolds, J.~B. Richards, H.~de~Wit, Acute-alcohol effects on the
  experiential discounting task ({E}{D}{T}) and a question-based measure of
  delay discounting, Pharmacology, Biochemistry and Behavior 83 (2006)
  194--202.

\bibitem{roblesa}
E.~Roblesa, P.~A. Vargas, R.~Bejarano, Within-subject differences in degree of
  delay discounting as a function of order of presentation of hypothetical cash
  rewards, Behavioural Processes 81 (2009) 260--263.

\bibitem{loew}
G.~F. Loewenstein, Frames of mind in intertemporal choice, Management Science
  34 (1988) 200--214.

\bibitem{murphy}
J.~G. Murphy, R.~E. Vuchinich, C.~A. Simpson, Delayed reward and cost
  discounting, The Psychological Record 51~(4) (2001) 571--588.

\bibitem{thaler}
R.~H. Thaler, Some empirical evidence on dynamic inconsistency, Economics
  Letters, 1981, vol. 8, issue 3, pages 201-207 8~(3) (1981) 201--207.

\bibitem{loew2}
G.~Loewenstein, D.~Prelec, Anomalies in intertemporal choice: Evidence and an
  interpretation, The Quarterly Journal of Economics 107~(2) (1992) 573--597.

\bibitem{patak}
M.~Patak, B.~Reynolds, Question-based assessments of delay discounting: do
  respondents spontaneously incorporate uncertainty into their valuations for
  delayed rewards?, Addictive Behaviors 32 (2007) 351–357 32 (2007) 351--357.

\bibitem{takahashi7}
T.~Takahashi, K.~Ikeda, T.~Hasegawa, A hyperbolic decay of subjective
  probability of obtaining delayed rewards, Behavioral and Brain Functions 2007
  3 (2007) 52.

\bibitem{samuelson}
P.~A. Samuelson, A note on measurement of utility, The Review of Economic
  Studies 4 (1937) 155--161.

\bibitem{rodriguez}
M.~L. Rodriguez, A.~W. Logue, Adjusting delay to reinforcement: comparing
  choice in pigeons and humans, Journal of Experimental Psychology: Animal
  Behavior Processes 14~(1) (1988) 105--117.

\bibitem{rachlin}
H.~Rachlin, A.~Raineri, D.~Cross, Subjective probability and delay, Journal of
  the Experimental Analysis of Behavior 55~(2) (1991) 233--244.

\bibitem{mcker}
T.~L. McKerchar, L.~Green, J.~Myerson, T.~S. Pickford, J.~C. Hill, S.~C. Stout,
  A comparison of four models of delay discounting in humans, Behavioural
  Processes 81 (2009) 256--259.

\bibitem{madden1}
G.~J. Madden, N.~M. Petry, G.~J. Badger, W.~K. Bickel, Impulsive and
  self-control choices in opioid-dependent patients and non-drug-using control
  participants: drug and monetary rewards, Experimental and Clinical
  Psychopharmacology 5~(3) (1997) 256--262.

\bibitem{kirb1}
K.~N. Kirby, N.~M. Petry, W.~K. Bickel, Heroin addicts have higher discount
  rates for delayed rewards than non-drug-using controls, Journal of
  Experimental Psychology: General 128~(1) (1999) 78--87.

\bibitem{prelec}
D.~Prelec, Decreasing impatience: a criterion for non-stationary time
  preference and ``hyperbolic'' discounting, Scandinavian Journal of Economics
  106~(3) (2004) 511--532.

\bibitem{rachlin1}
H.~Rachlin, Notes on discounting, Journal of the Experimental Analysis of
  Behavior 85~(3) (2006) 425--435.

\bibitem{myerson}
J.~Myerson, L.~Green, Discounting of delayed rewards:models of individual
  choice, Journal of the Experimental Analysis of Behavior 64~(3) (1995)
  263--276.

\bibitem{simpson}
C.~A. Simpson, R.~E. Vuchinich, Reliability of a measure of temporal
  discounting, The Psychological Record 50 (2000) 3--16.

\bibitem{cajueiro}
D.~O. Cajueiro, A note on the relevance of the $q$-exponential function in the
  contextof intertemporal choices, Physica A 364 (2006) 385--388.

\bibitem{tiago}
T.~J. Arruda, R.~S. González, C.~A.~S. Terçariol, A.~S. Martinez, Arithmetical
  and geometrical means of generalized logarithmic and exponential functions:
  generalized sum and product operators, Physics Letters A 372 (2008)
  2578--2582.

\bibitem{martinez}
A.~S. Martinez, R.~S. Gonz\'alez, A.~L. Esp\'indola, Generalized exponential
  function and discrete growth models, Physica A 388 (2009) 2922--2930.

\bibitem{martinez1}
A.~S. Martinez, R.~S. Gonz\'alez, C.~A.~S. Ter{\c c}ariol, Continuous growth
  models in terms of generalized logarithm and exponential functions, Physica A
  387 (2008) 5679--5687.

\bibitem{2010arXiv1010.2950S}
A.~S. {Martinez}, B.~C.~T. {Cabella}, F.~{Ribeiro}, {Scaling Function,
  Universality and Analytical Solutions of Generalized One-Species Population
  Dynamics Models}, ArXiv e-prints 1010.2950.

\bibitem{martinez2}
C.~Anteneodo, C.~Tsallis, A.~S. Martinez, Risk aversion in trade transactions,
  Europhysics Letters 59~(5) (2002) 635--641.

\bibitem{martinez3}
A.~S. {Martinez}, R.~S. González, C.~A.~S. Terçariol, { ?Generalized
  Probability Functions}, Advances in Mathematical Physics 2009 (2009) 206176.

\bibitem{takahashi2}
T.~Takahashi, A comparison between tsalliss statistics-based and generalized
  quasi-hyperbolic discount models in humans, Physica A 387 (2008) 551--556.

\bibitem{takahashi3}
T.~Takahashi, A comparison of intertemporal choices for oneself versus someone
  else based on tsallis statistics, Physica A 385~(2) (2007) 637--644.

\bibitem{takahashi4}
T.~Takahashi, A probabilistic choice model based on tsallis statistics, Physica
  A 386~(1) (2007) 335--338.

\bibitem{stevens}
S.~S. Stevens, On the psychophysical law, The Psychological Review 64~(3)
  (1957) 153--181.

\bibitem{wong}
K.~H. Norwich, W.~Wong, Unification of psychophysical phenomena:the complete
  form of {F}echner's law, Perception \& Psychophysics 59~(6) (1997) 929--940.

\end{thebibliography}

\end{document}